\documentstyle[npb,twoside,fleqn,epsfig]{article}
\newcommand{\half}{{\textstyle\frac{1}{2}}}
\newcommand{\as}{\alpha_s}
\newcommand{\das}{\delta\as}

\newcommand{\eps}{\epsilon}

\newcommand{\beq}{\begin{equation}}
\newcommand{\eeq}{\end{equation}}

\newcommand{\cl}[1]{{\cal #1}}
\newcommand{\fhat}{\hat{\cl{F}}}

\newcommand{\AmS}{{\protect\the\textfont2
  A\kern-.1667em\lower.5ex\hbox{M}\kern-.125emS}}

\title{\begin{flushright}\normalsize
  \vspace{-1cm}
  Bicocca--FT--99--15\\
Cavendish-HEP-99/06\\
hep-ph/9906236
  \vspace{0.5cm}
\end{flushright}
$1/Q^2$ corrections in DIS fragmentation functions}

\author{M.~Dasgupta\address{INFN-Sezione di Milano, Via Celoria 16, \\
        Milano 20133, Italy.}
\thanks{Talk presented at 7th International Workshop on Deep Inelastic Scattering and QCD, Zeuthen, Germany, April 1999.}        
and
G.E.~Smye\address{Cavendish Laboratory, University of Cambridge, \\
        Madingley Road, Cambridge, CB3 0HE, United Kingdom.}
        \thanks{Research supported by the UK Particle Physics and Astronomy
                Research Council.}}
\begin{document}
 
\begin{abstract}
We present results of recent calculations \cite{nonsing,smyetwo} of power corrections to single-hadron inclusive momentum distributions in the current hemisphere of the Breit frame in DIS. Though results are presented for both singlet and non-singlet scattering contributions, we emphasise the fact that the former is dominant in the kinematical region studied at HERA. 
\end{abstract}
\maketitle
\section{INTRODUCTION}
The work described in the following article is largely motivated by the recent experimental interest in the study of hadronic final state momentum distributions in the current hemisphere of the Breit frame of DIS. The observable in question is similar to the hadronic energy distribution in $e^{+}e^{-}$ annihilation and is related in an analogous manner to the universal parton to hadron fragmentation functions. However in order to compare theoretical predictions for these DIS momentum distributions accurately with the HERA data, we must take into account the existence of $1/Q^2$ non-perturbative effects (power corrections) from both the singlet and non-singlet processes.  

We wish to study the distribution $F^h(z)$, defined by
\begin{equation}
\label{fragdef}
F^h(z;x,Q^2) = \frac{d^3\sigma^h}{dxdQ^2dz}\left/\frac{d^2\sigma}{dxdQ^2}\right.\; ,
\end{equation}
for a given hadron species $h$, as a function of the variable $z = 2 p_h\cdot q/q^2$, where $p_h$ is the four-momentum of the resultant hadron and $q$ the photon four-momentum. Since the fragmentation products of the remnant of the nucleon are expected to be travelling in directions close to that of the incoming nucleon, i.e. in the `remnant hemisphere' $p_z \le 0$, we consider only hadrons produced in the `current hemisphere' $p_z \ge 0$. Such hadrons are expected to be fragmentation products of the scattered parton. Thus $z$ takes values $0\le z\le 1$. The other quantities $x$, $Q^2$ have their usual DIS meanings. 

One has to consider power corrections to both the numerator and the denominator of Eq.~(\ref{fragdef}) each of which have a leading power correction that goes as $1/Q^2$. The denominator is just the familiar DIS cross section which can be decomposed in terms of transverse and longitudinal structure functions for which the results are presented in Refs.~\cite{A2val} and \cite{smyeone}. For the numerator one has an identical decomposition in terms of generalised ($z$ dependent) structure functions to which we also compute power corrections. We present here, results for both singlet and non-singlet contributions to $F^h(z;x,Q^2)$. The calculations described here were done using the dispersive technique which we briefly comment on below.
\section{CALCULATIONS AND RESULTS}
The calculation for the non-singlet piece follows the usual dispersive method first described in \cite{dmw}. In this method power corrections correspond to non-analytic pieces in a fake gluon mass $\epsilon$, in the ${\mathcal{O}}(\alpha_s)$ calculation of a given observable, represented by the characteristic function $\mathcal{F}(\epsilon)$. Importantly, the normalisation of the result is in terms of  hopefully universal parameters, which turn out to be various moments of a  universal strong coupling $\as(\mu^2)$, defined for all scales $\mu^2$.

In order to calculate power-behaved contributions to QCD observables involving more than a single renormalon chain, like the singlet sector here, we need to generalise the standard dispersive treatment, the details of which can be found in references \cite{smyetwo,smyeone}. 
Here we just mention that in cases involving for instance two gluons, like the forward scattering amplitude of the DIS singlet graphs, we need to introduce two dispersive variables (gluon masses) and our characteristic function ${\mathcal{F}}(\epsilon_1,\epsilon_2)$ is now an ${\mathcal{O}}(\alpha_s^2)$ quantity. However it turns out that the answer can  be expressed in terms of functions  $\fhat$ which are obtained from the characteristic function by setting one of the gluon masses to zero while retaining the other. Hence it is still the non-analytic terms in a single variable $\epsilon$ which correspond to power corrections. 

In the present case  (DIS fragmentation functions) we find the following types of non-analyticity which translate into $1/Q^2$ corrections as below: 
\beq
\fhat\sim c\eps\log\eps
\;\;\Longrightarrow\;\;
\delta F = c\frac{D_1}{Q^2}
\eeq
and
\beq
\fhat\sim \half c^\prime\eps\log^2\eps
\;\;\Longrightarrow\;\;
\delta F = c^\prime\frac{D_1}{Q^2}\log\frac{D_2}{Q^2}\;,
\eeq
where $D_1$ and $D_2$ are defined by
\beq
       D_1 \equiv \int_0^\infty d\mu^2\left[2\as\das-(\das)^2\right]
\eeq
\beq
\log D_2    \equiv \frac{1}{D_1}\int_0^\infty d\mu^2\log\mu^2\left[2\as\das-(\das)^2\right]\;.
\eeq
and
$$\das(\mu^2)=\alpha_s(\mu^2)-\alpha_s^{{\mbox{\tiny{PT}}}}(\mu^2)$$
represents a non-perturbative modification to the standard perturbative form of thr coupling. We emphasise the fact that in the above formula $\alpha_s^{{\mbox{\tiny{PT}}}}$ stands for the perturbative expansion of $\alpha_s(\mu^2)$ in terms of $\alpha_s(Q^2)$, where this expansion is truncated at whatever order the perturbative result for a given observable has been computed. Hence in most cases of current interest this would be at ${\mathcal{O}}(\alpha_s^2)$.

We calculate the singlet and non-singlet contributions to the fragmentation functions, retaining terms up to $\cl{O}(\eps)$ non-analytic at $\eps = 0$. We find terms that diverge as $\log\eps$ and $\log^2\eps$ arising from collinear splitting; the gluon mass here behaves as a regulator, and the divergence is factored into the scale dependence of the parton distributions.

\begin{figure}[htb]
\epsfig{file=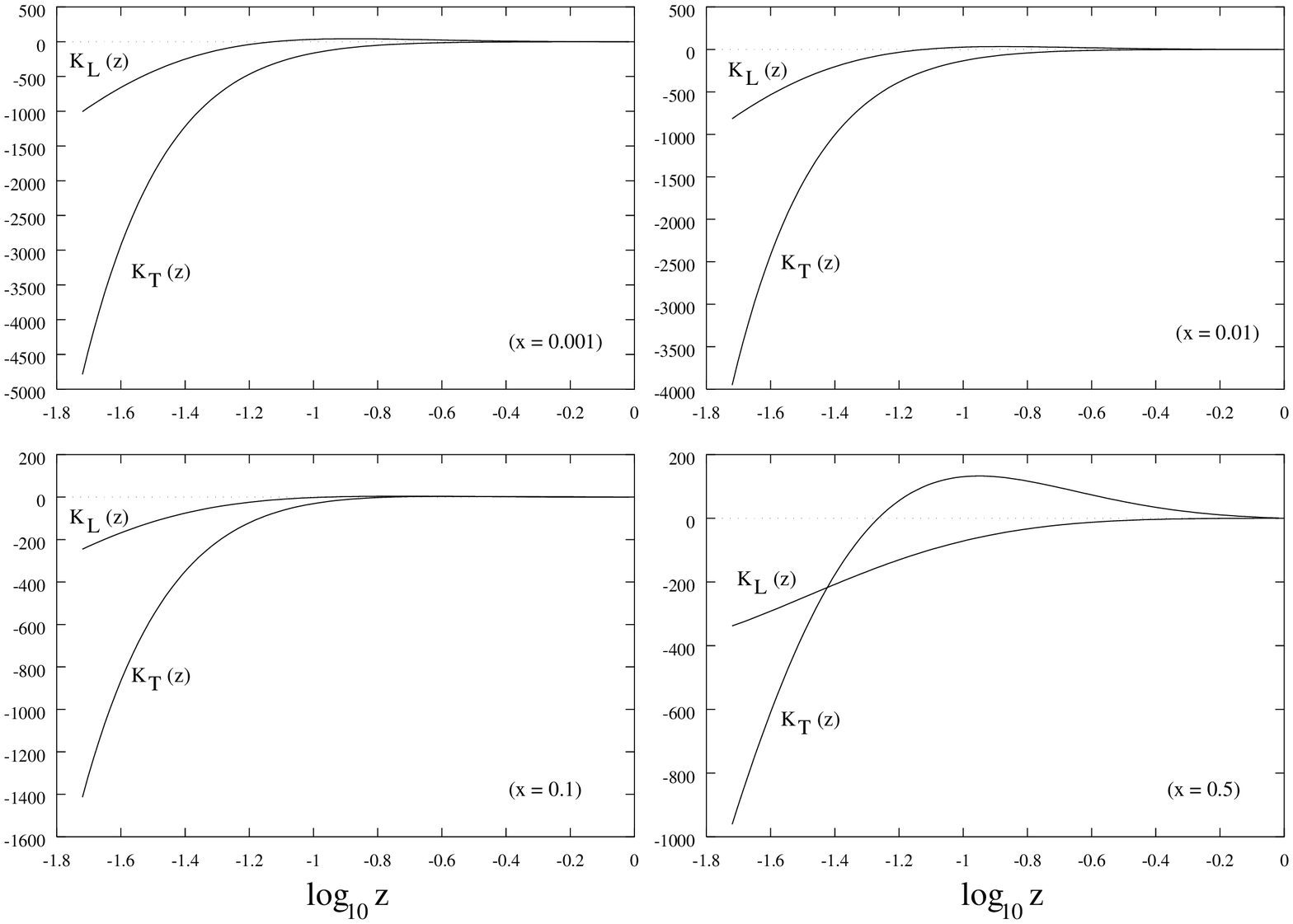,height=55mm,width=75mm}
\vspace{-1.5cm}
\caption{\label{singgr}Graphs showing $K_T(z)$ and $K_L(z)$ for singlet fragmentation at fixed values of $x$.}
\vspace{-0.5cm}
\end{figure}

The $1/Q^2$ corrections arise from terms proportional to $\eps\log\eps$ and $\eps\log^2\eps$. In the singlet case we obtain the correction
\beq
\delta F^h = \frac{D_1}{Q^2}\frac{T_R C_F}{(2\pi)^2}\left[K_T+\frac{2(1-y)}{1+(1-y)^2}K_L\right]
\eeq
where $K_T(z,x)$ and $K_L(z,x)$ are shown in figure \ref{singgr} and $y$ is the usual DIS variable. Note that if $D_1$ is positive, we would expect to see a large negative power correction at small $z$.

\begin{figure}[htb]
\epsfig{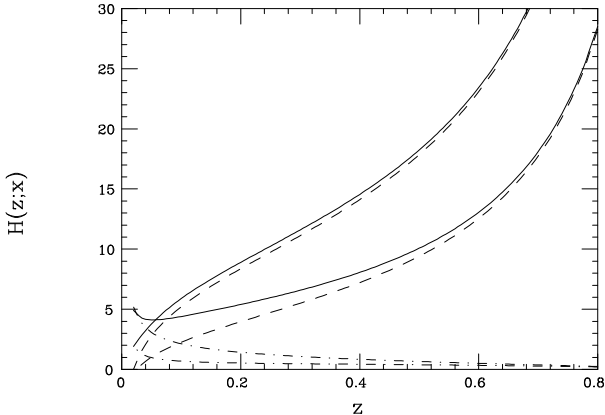}
\vspace{-1.5cm}
\caption{\label{nonsgr}Values of $H(z,x)$ for non-singlet fragmentation. Dashed, dot-dashed and solid curves are quark, gluon and total fragmentation; the two sets of curves are for $x=0.3$ (upper) and $0.1$ (lower).}
\vspace{-1cm}
\end{figure}

In non-singlet scattering we find the relative correction
\beq
\frac{\delta F^h}{F^h} = \frac{\cl{A}_2}{Q^2}H(z,x)
\eeq
where $H(z,x)$ is shown by figure \ref{nonsgr}. This gives a large positive correction at large $z$, but only a small correction at small $z$.

The detailed calculations and algebraic expressions for results can be found in \cite{nonsing,smyetwo}.

\section{CONCLUSIONS AND PROSPECTS}
The results presented in this article clearly indicate that at small values of the fragmentation variable $z$, one would expect the singlet contribution to be dominant. Moreover at small values of the Bjorken variable $x$, we would anyhow expect the singlet scattering mechanism to start becoming important due to the rise in the gluon density.  

The parameter $D_1$ that appears in the singlet result has to be phenomenologically determined (extracted once and for all from the data). Results are relatively insensitive to the other parameter $D_2$. Unfortunately however, before one can compare the predictions made here with the experimental data one needs to clarify several issues. 

From a technical viewpoint one has to learn how to treat the singlet piece as a gluonic contribution rather than an ${\mathcal{O}}(\alpha_s^2)$ non-singlet contribution. In other words, the result presented here was obtained by treating the incoming gluon (that scatters via $q \bar{q}$ production off the photon) as an internal line radiated off an incoming quark. This is the only way to obtain the power correction. However to reinterpret this as a singlet contribution we have to find a meaningful way to factorise off the quark to gluon splitting and then convolute with the gluon density rather than the quark density. 

Then there is also the question of doing improved perturbative calculations. At present the NLO program CYCLOPS \cite{Graud} is used to obtain the perturbative prediction. However in actual fact, at small values of $z$ we would expect terms that vary as $\log(1/z)$ to spoil the convergence of the perturbative series. Hence the need arises for resummed perturbative estimates which take these terms into account at all orders. 

Another  relatively minor consideration is that for the present results we calculate the distribution differential in $z$, the hadron longitudinal momentum fraction, rather than the energy fraction, since it turns out to be complicated to analytically compute the latter in DIS. Hence when comparisons are made with data one must make sure that we are using the same fragmentation variable although qualitative features can be expected to be similar for both choices. 

In fact at this conference we saw experimental results for the energy distributions where the deviation from NLO perturbation theory was indeed large and negative as we expect at small $z$ \cite{Kant}. Remarkably the data can be extremely well described using a simple power correction ansatz of Dokshitzer and Webber also presented here. However the ansatz is not a theoretical prediction in any strict sense but rather an educated guess at the shape of the correction. Hence for a full theoretical picture we need to understand the problems mentioned above. Once this is done we can proceed with extracting $D_1$ from the data and using it in fits to other singlet observables.


\begin{thebibliography}{9}
\bibitem{nonsing} M. Dasgupta, G.E. Smye and B.R. Webber, JHEP 04 (1998) 017.
\bibitem{smyetwo}    G.E. Smye, hep-ph/9812251 (to be published in Nucl. Phys. B).
\bibitem{A2val}   M. Dasgupta and B.R. Webber, Phys. Lett. B 382 (1996) 273.
\bibitem{smyeone}   G.E. Smye, Nucl. Phys. B 546 (1999) 315.
\bibitem{dmw}     Yu.L. Dokshitzer, G. Marchesini and B.R. Webber, Nucl. Phys. B 469 (1996) 93.
\bibitem{Graud} D.~Graudenz, CYCLOPS program.
\bibitem{Kant} D.~Kant, these proceedings.

\end{thebibliography}
\end{document}